# Single Microhole per Pixel in CMOS Image Sensor with Enhanced Optical Sensitivity in Near-Infrared

E. Ponizovskaya Devine[1,*], Wayesh Qarony[2], Ahasan Ahamed[2], Ahmed S Mayet[2], Soroush Ghandiparsi[2], Cesar Bartolo-Perez[2], Aly F Elrefaie[1], Toshishige Yamada[1,3], Shih-Yuan Wang[1], M. Saif Islam[2,*]

[1]W&WSens Devices, Inc., 4546 El Camino, Suite 215, Los Altos, California 94022, USA

[2]Electrical and Computer Engineering, University of California—Davis, Davis, California 95618, USA

[3]Electrical Engineering, Baskin School of Engineering, University of California, Santa Cruz, California 95064, USA

*Corresponding authors: eponizovskayadevine@ucdavis.edu

sislam@ucdavis.edu (M. Saif Islam)

**Abstract**

Silicon photodiode-based CMOS sensors with backside-illumination for 300–1000 nm wavelength range were studied. We showed that a single hole in the photodiode increases the optical efficiency of the pixel. In near-infrared wavelengths, the enhancement allows 70% absorption in a 3 µm thick Si. It is 4x better than for the flat pixel. We compared different shapes and sizes of single holes and holes arrays. We have shown that a certain size and shape in single holes pronounce better optical efficiency enhancement. The crosstalk was successfully reduced with trenches between pixels. We optimized the trenches to achieve minimal pixel separation for 1.12 µm pixel.

## 1. Introduction

The interest in the complementary metal-oxide-semiconductor (CMOS) image sensors is increasing due to the growing demand for mobile imaging, digital cameras, surveillance, monitoring, and biometrics. Image sensors have been adopted in a great number of products [1-3]. The applications, such as monitoring and surveillance demand high sensitivity options in near-infrared for night vision application. We are considering a stacked structure similar to the one reported by Sony that exhibited low noise [4] and were back-side illuminated and layered over a chip where signal processing circuits are formed. One advantage of this configuration is that large-scale circuits can be mounted on a relatively small chip. Since each section is formed on a separate chip, specialized manufacturing processes can be used to produce high performance imaging sensors and a state-of-the-art circuit section, enabling higher resolution, multi functionality, and a compact size. Besides chip size, the stacking technology improves dark characteristics, such as noise and white pixels, by optimizing the sensor process independently.

Two different methods are implemented for near-infrared imaging: (i) application of a separate near-infrared filter, based on plasmonic techniques [5], or gratings [6]; and (ii) standard low-cost pigment filters [7] that are transparent in near-infrared wavelengths. Infrared images are reconstructed based on the techniques discussed in Ref. [8]. To increase the quantum efficiency, the surface of the pixels is textured with an array of small inverted pyramids with dimensions in the range of 400 nm [9]. We model a Bayer array with a commonly used color filter array [10].

In general, CMOS image sensors can have pixel sizes of 2 µm × 2 µm or even smaller dimensions [11]. Such small pixels contribute to a major challenge in suppressing the parasitic charge exchanges between neighboring pixels (crosstalk). Although the inverted pyramid arrays increase the crosstalk, it was reduced using the deep trench isolation (DTI) [12,13].

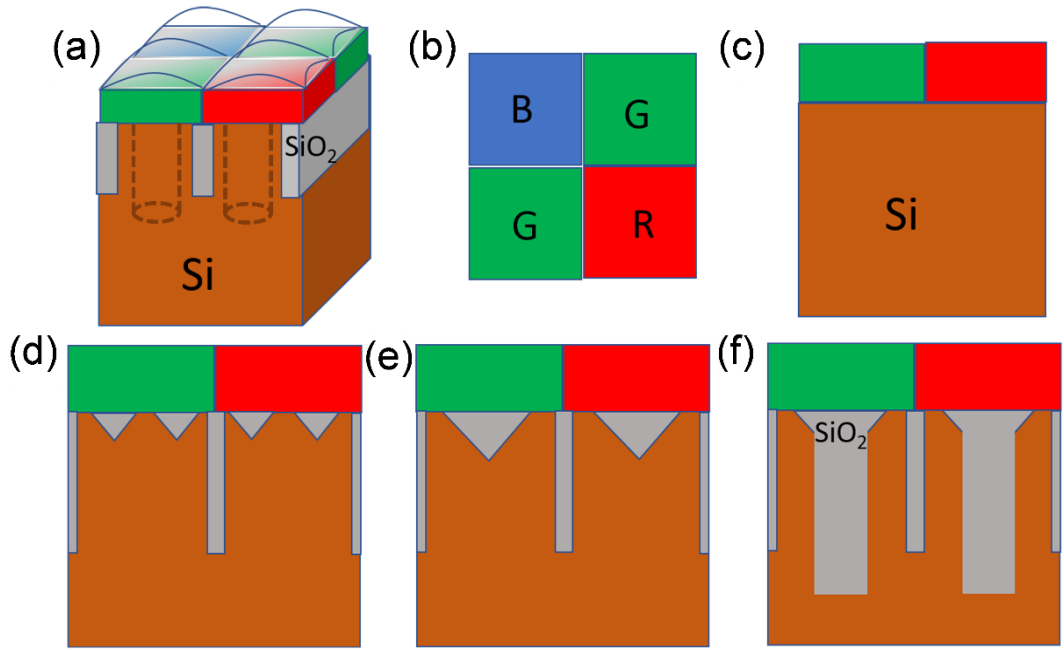

**Fig. 1**. Schematic diagram of CMOS image pixels: a) view of the pixel with micro lenses, color filters, and trenches; b) Bayer filter array c) planar pixel with color filters d) pixels with inverted pyramids array; e) pixel with a single inverted pyramid, d) pixel with single funnel holes.

We discuss the different approaches to increase the optical efficiency in infrared wavelengths. It was shown [14] that the microholes arrays can drastically increase the optical efficiency of Si in 800-1000 nm wavelength range. Here, we propose to use a single hole per diode. The hole size is in the range of 800-1000 nm. We simulated CMOS image sensors with different hole shapes, such as pyramid, cylindrical, and funnel shapes [15], and we compared the optical efficiency and the crosstalk for each shape. The deep trench structure [12,13] with Si-SiO2 interface is used as a barrier against electron diffusion. This also helps in confining light within the pixels by acting as a reflector between pixels. We simulated 1.12 µm wide and 3 µm deep pixels with trenches of 0 to 250 nm width and 0 to 2.9 µm depth. Our simulations show that a single funnel hole in a device provides better efficiency compared to an array. A funnel shaped hole with the size comparable to the wavelength converts propagation direction of light into lateral, and provides better light trapping effect and reduce the reflection, as was shown in Ref [16,17, 21] that implemented this technique for high-speed Si photodetectors The same shape single holes were used for the simulations. We used the Lumerical FDTD software and the methodology of the simulation that was described in [1] to calculate the absorption in each pixel. The single funnel hole enhanced the absorption by more than 60% at 850 nm and 940 nm wavelength for pixel of 1.12×1.12 μm lateral dimension and 3 μm depth. This is two times thinner than the devices reported in [1] and significantly exceeds the results reported in [2-3] for inverted pyramids arrays.

## 2. Optical Simulation Methodology

We use Finite Difference Time Domain (FDTD) [18] Lumerical software [19] to solve Maxwell's curl equations for the unit cell of Bayer array (Fig.1), where Bloch boundary conditions around the cell and Perfectly Matched Layer (PML) in the direction normal to the surface were considered. CMOS image sensor model includes lenses, red, green, and blue filters with a thickness of 900 nm, antireflection coating, and a 3 μm thick Si with SOI substrate. Micro-lens with a radius and thickness of 1 μm and 500 nm was also assumed on the top of each sensor. Additionally, different shapes and sizes of holes were considered as a photon-trapping structure to enhance the absorption efficiency or optical efficiency of the sensor, where trenches filled with silicon oxide were assumed. Herein, we compared several designs: flat pixel with no surface photon-trapping structure (Fig.1c), an array of holes with 400 × 400 nm inverted pyramids [9] (Fig.1d), pixel with a single inverted pyramid of 900 × 900 nm (Fig1.e), a funnel hole with a diameter of 900 nm and an angle of 60 degree wall that merges with 800 nm diameter and 2 μm deep cylindrical hole (Fig.3f), and a cylindrical hole with a diameter and depth of 800 nm and 2 μm, respectively.

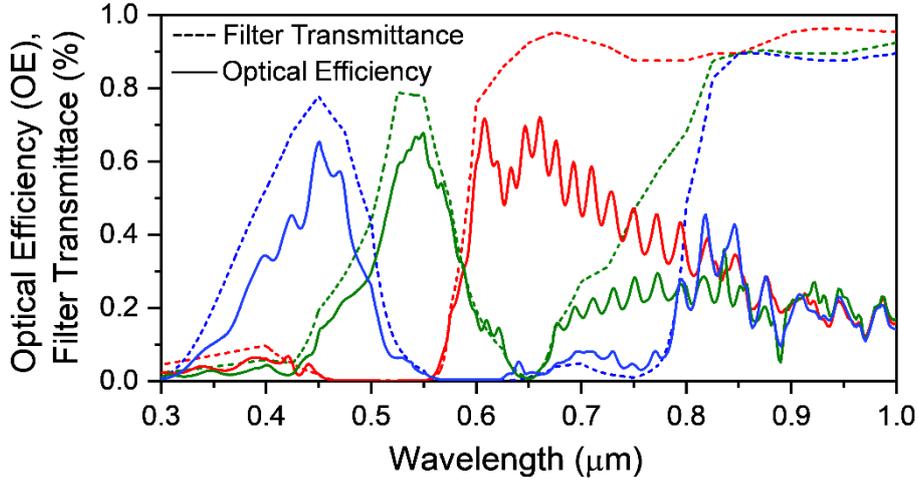

**Fig. 2**. Simulated transmittance after filters (dashed lines) and optical efficiency (OE) of a flat pixel (solid lines).

A plane wave source was considered with normal incidence to the surface for the wavelengths ranging between 300 and 1100 nm. The Poynting vector (P) was measured around the cells.

For a given current J in the pixel and electric field, E, the energy absorbed in volume V is calculated by applying the divergence theorem for stable state as below:

$$\int_V \frac{1}{2} \text{Re}(E \cdot J^*) \cdot dv = \int_V \frac{1}{2} \nabla \cdot \text{Re}(E \times H^*) \cdot dv$$
$$= -\oint_S \frac{1}{2} \left[ \text{Re}(E \times H^*) \cdot \vec{n} \right] ds \tag{1}$$

where, S is the surface that surrounds the volume V and $\vec{n}$ is the unit vector normal to the surface S. Thus, to calculate the optical absorption in each pixel power, we integrate the Poynting vector normal to the surface over the surface of the depletion region of the pixel. The results present the difference between the real parts of the Poynting flux entering the volume at the surface between the filters and the pixel ($P_{in}$

= $Re(E \times H^*)$), and the Poynting flux leaving the pixel ($P_{out}$), that are simulated with FDTD method. The optical efficiency (OE) is calculated as following:

$$OE = \frac{P_{in} - P_{out}}{P_{inc}} \qquad (2)$$

where, $P_{inc}$ is the Poynting vector of the incident light calculated above the lenses and filters. In this study, it is assumed that the quantum efficiency (QE) is proportional to the optical efficiency (OE) [20]. The filter's spectrum is determined by the real and imaginary parts of the refractive index n and k (table in the supplement). They were recalculated from the transmission for the pigment filters of 900 nm thickness as reported in Ref. [7].

The transmission profile of flat and photon-trapping device is shown in Fig. 2. Compared to the flat device (solid lines), the devices with microhole structures (dashed line) exhibits higher transmission. The maximum possible optical efficiency for the blue, red, and green filters are calculated to be $OE_{blue}$= 80% at 440 nm, $OE_{green}$ = 85% at 550 nm, and $OE_{red}$ = 85% at 650 nm, whereas the filters are transparent in the near-infrared. However, the absorption efficiency of Si is very week in the near-infrared, where light trapping strategies are required to enhance the light absorption by optical light trapping, bending, or slowing down the light. In this case, nano/micro holes are used to improve the absorption efficiency (see Fig.2 dashed curve).

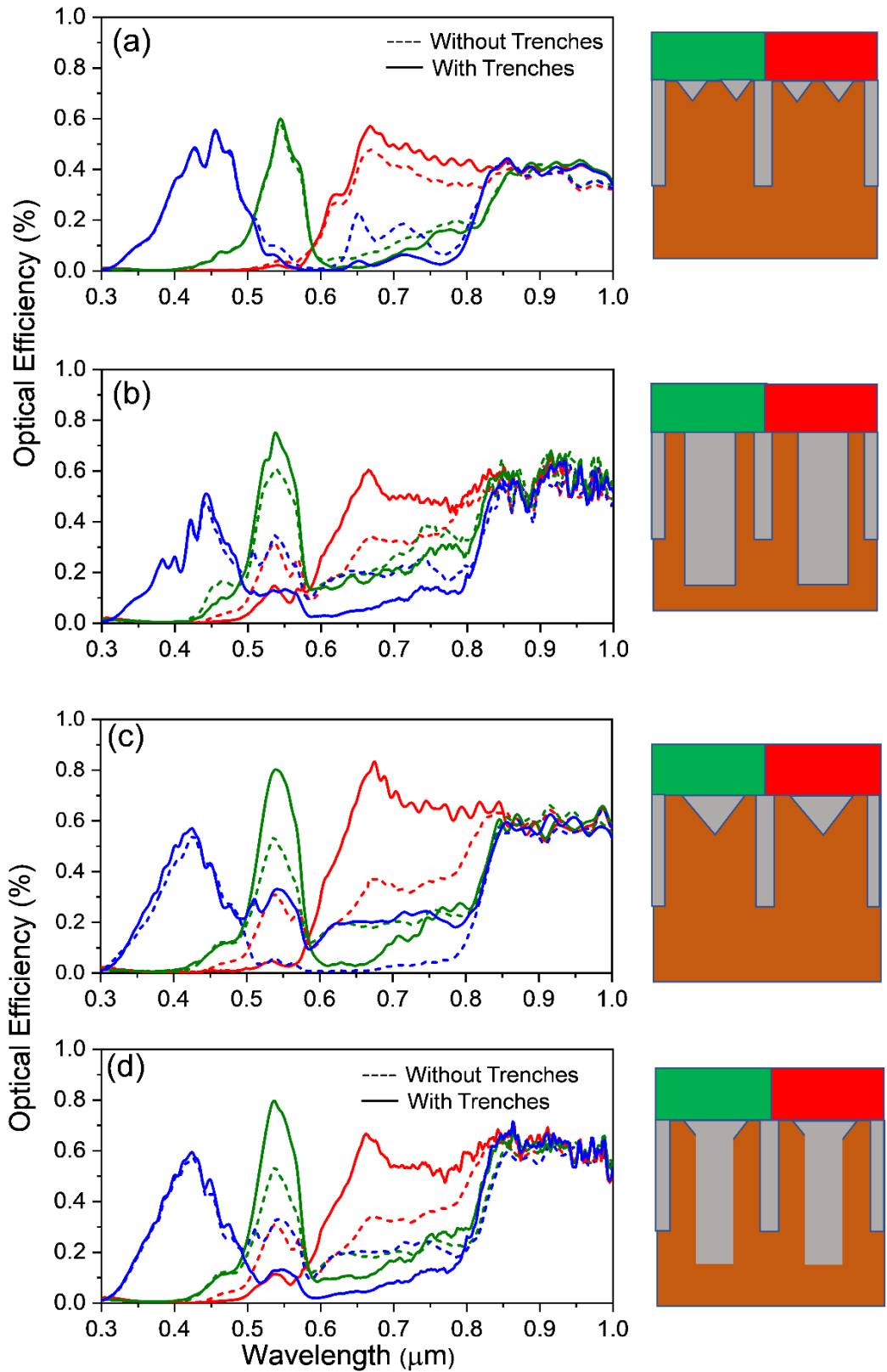

**Fig. 3.** Calculated optical efficiency (OE) with (solid) and without (dashed) trenches for (a) inverted pyramids array, (b) single cylindrical holes, (c) single inverted pyramid, and (d) funnel single hole.

**Table I:** Crosstalk of the simulated image sensors integrated with photon-trapping holes with inverted pyramids array, single cylindrical pyramid, single cylindrical hole, and single funnel hole, where a flat pixel is simulated for as a reference.

| Holes Design/Structure | Pixel Color | Crosstalk index No DTI/DTI | OE(%)@ 850/940 nm |
|---|---|---|---|
| Flat Pixel | Red-Green | 4 | 22/15 |
| | Green-Red | 20 | |
| | Blue-Red | 24 | |
| Inverted Pyramid Array | Red-Green | 6.1/26 | 35/20 |
| | Green-Red | 13.5/30 | |
| | Blue-Red | 24/24 | |
| Single Inverted Pyramid | Red-Green | 1.6/14.7 | 60/59 |
| | Green-Red | 2.8/20 | |
| | Blue-Red | 50/24 | |
| Single Cylindrical Pyramid | Red-Green | 1.5/2.9 | 58/62 |
| | Green-Red | 2.1/5.6 | |
| | Blue-Red | 76/26 | |
| Single Funnel Pyramid | Red-Green | 76/26 | 71/62 |
| | Green-Red | 1.34/69 | |
| | Blue-Red | 1.68/7.9 | |

### 3. Results and Discussion

First, simulations were performed for the planar sensors (Fig.2). Dashed curves show the transmittance after the filters and the solid curve shows the OE in the pixels under the filter with the corresponding color. For this case, there is relatively low crosstalk for vertical illumination. As expected for silicon-based photodiodes sensors, the OE in infrared wavelengths is much lower than the OE for the visible wavelengths of blue, red, and green. The introduction of array of small inverted pyramids on the surface of an optical sensor can increase the optical efficiency as well as the crosstalk in the infrared wavelengths [8].

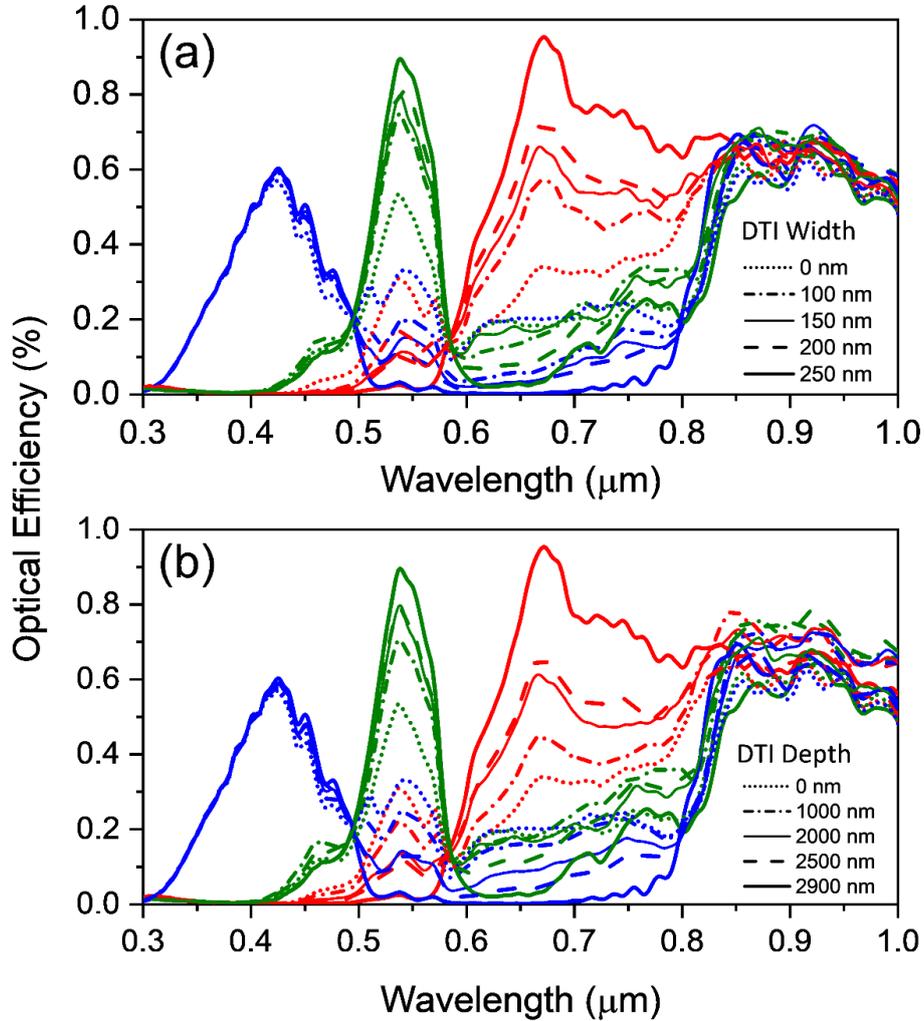

**Fig. 4.** Optimization of DTI (a) width and (b) depth in the CMOS image sensor. The influence of DTI width and depth was investigated in optical efficiency by varying them from 0 nm to 250 nm and 0 nm to 2900 nm, respectively.

Fig. 3 compares the OE of the inverted pyramids of 400×400 nm array [9] (a), cylindrical (b), single inverted pyramid 900×900 nm (c), and funnel hole (d). Solid lines represent the pixel with trenches of 250 nm width and 2.5 μm deep, whereas the dashed line shows the pixels without trenches. While the hole arrays increase the OE at NIR to about 40%, a single inverted pyramid hole of 900×900 nm can increase the OE to more than 60% and an optimized single funnel shaped hole shows an OE up to 70%. All the microstructures increased the absorption in the pixels for blue, green, and red, with higher enhancement in the last two. However, the integration of holes in the devices can increase the crosstalk between pixels, but it can be effectively be reduced by the implementation of trenches without any decrease in OE.

Table 1 shows the crosstalk with and without trenches for all the shapes studied. The trenches could be optimized to provide the smallest crosstalk by varying their depths and thicknesses. Fig.4 shows the effect of the variation of the trenches size. As we can see, the crosstalk decreases with the depth of the trench up to 2.9 µm and with increasing the width up to 250 nm. However, a reasonable crosstalk could be achieved with a depth of 2.5 µm and width even up to 150 nm. The crosstalk for a blue pixel from a red pixel is defined as a ratio between the response of the blue pixel at the 440 nm wavelength to the response of the red pixel at the same wavelength. In the case of NIR wavelengths, there will be an equal response from all three pixels and will be represented as a grayscale picture. The simulations show that while the crosstalk decreased, the OE is increased for all the colors due to the use of single holes with trenches.

**Conclusion**

We presented silicon CMOS sensors with backside-illumination designed using 3 µm thick silicon and with 1.12 µm × 1.12 µm pixel size in RGB. Optical simulations were conducted using FDTD methods. We have shown that a single hole with a size of about 900 nm increases the optical efficiency (OE) of the image sensors for all the colors and in near-infrared wavelengths. In the NIR wavelengths, the OE of the sensors could be as high as 70%. The OE achieved in the funnel shape hole, was much higher than the one with inverted pyramids arrays that were reported before. The crosstalk increased due to the use of holes, but it was reduced back to normal level with the implementation of trenches. We optimized trenches with 150 nm width and 2.5 µm depth for minimum crosstalk.

**Acknowledgment**

We acknowledge W&WSens Devices for financial support.

**Appendix**

The parameters of the filters that were used in the model were taken from [6]. Table for the filters parameter that was used in the simulations shows the n and k for the refractive index represented as n+ik.

**Table II**: Filters Parameters

| Wavelength, nm | Blue | | Green | | Red | |
|---|---|---|---|---|---|---|
| | n | k | n | k | n | k |
| 300 | 1.55 | 0.055 | 1.62 | 0.4 | 1.54 | 0.415 |
| 400 | 1.54 | 0.07 | 1.6 | 0.5 | 1.54 | 0.325 |
| 500 | 1.54 | 0.215 | 1.58 | 0.405 | 1.53 | 0.155 |
| 600 | 1.54 | 0.8 | 1.57 | 0.05 | 1.53 | 0.048 |
| 700 | 1.53 | 0.45 | 1.57 | 0.11 | 1.52 | 0.033 |
| 800 | 1.53 | 0.455 | 1.57 | 0.105 | 1.52 | 0.375 |
| 900 | 1.52 | 0.465 | 1.56 | 0.09 | 1.52 | 0.0185 |
| 100 | 1.52 | 0.39 | 1.56 | 0.055 | 1.52 | 0.015 |